\documentclass[a4paper]{article}

\usepackage{INTERSPEECH2021}
\usepackage{fvextra}

\title{Selective Pseudo-labeling and Class-wise Discriminative Fusion for Sound Event Detection}

\name{Yunhao Liang$^1$, Yanhua Long$^1$\thanks{Yanhua Long is the corresponding author. The work is supported by the National Natural Science Foundation of China (Grant No.62071302).}, Yijie Li$^2$, Jiaen Liang$^2$}
\address{
  $^1$Key Innovation Group of Digital Humanities Resource and Research, \\
  Shanghai Normal University, Shanghai, China\\
  $^2$Unisound AI Technology Co., Ltd., Beijing, China}
\email{winnerahao@163.com, yanhua@shnu.edu.cn, \{liyijie,liangjiaen\}@unisound.com}

\begin{document}

\maketitle
\begin{abstract}

In recent years, exploring effective sound separation (SSep) techniques to
improve overlapping sound event detection (SED) attracts more and more attention.
Creating accurate separation signals to avoid the catastrophic error
accumulation during SED model training is very important and challenging.
In this study, we first propose a novel selective
pseudo-labeling approach, termed SPL, to produce high confidence separated
target events from blind sound separation outputs. These target events are then
used to fine-tune the original SED model that pre-trained on
the sound mixtures in a multi-objective learning style. Then,
to further leverage the SSep outputs, a class-wise discriminative fusion
is proposed to improve the final SED performances,
by combining multiple frame-level event predictions of both sound mixtures and their 
separated signals. All experiments are performed on the public DCASE 2021 Task 4 dataset,
and results show that our approaches significantly outperforms the official
baseline, the collar-based $F1$, PSDS1 and PSDS2 performances
are improved from 44.3\%, 37.3\% and 54.9\% to 46.5\%, 44.5\% and 75.4\%, respectively.

\end{abstract}

\section{Introduction}
\label{sec:Intro}

Sound conveys a wide range of important information in our daily lives.
These sounds make us better to understand changes in our physical environment
and to perceive events occurring around us.
Sound event detection (SED) is a task of detecting both the onset and offset
of a sound event. It has widespread applications for real-world intelligent human-interaction
systems, including smart home devices, mobile devices, smart headphones, etc
~\cite{southern2017sounding, radhakrishnan2005audio}.
However, there are many challenges in real SED applications, including audio 
signal degradation from acoustic reverberation~\cite{benetos2016detection}, or
the real recordings may contain
not only overlapping predefined target events, but also non-target events and
a large number of environmental noise. These overlapping recordings are normally called
sound mixtures in the sound separation (SSep) field,
their complicated acoustic nature brings more interference to
the distinction of target event categories and the detection of event time-stamps.

In the past few years, as one of the most important challenge in SED tasks,
the overlapping sound events problem has been tackled in many perspectives.
In~\cite{salamon2015feature, serizel2020sound}, authors trained the SED model
as a multi-label system where the most energetic sound events are usually detected
more accurately than the rest. While in ~\cite{mesaros2016tut},
a set of binary classifiers were built to alleviate the SED
overlapping problem. Other approaches based on signal processing
techniques were also investigated, such as
using factorization techniques on the inputs of the classifier~\cite{benetos2016detection, bisot2017overlapping},
or exploiting spatial information when available~\cite{adavanne2018multichannel}, etc.

\begin{figure*}[th]
  \centering
  \includegraphics[width=15cm]{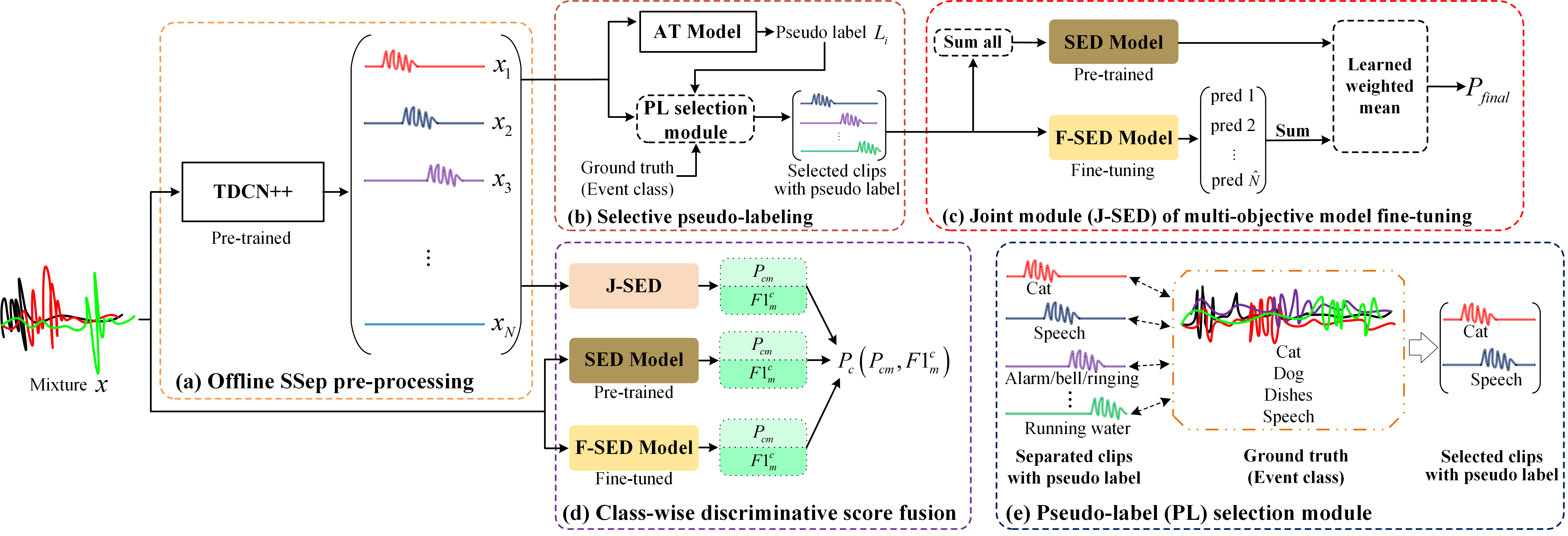}\\
  \caption{System architecture of the proposed SSep-SED joint framework.}
  \label{fig:joint training framework}
\end{figure*}

In recent years, with the Task 4 Challenge of Detection and
Classification of Acoustic Scenes and Events (DCASE) that launched
in 2020 and 2021 \cite{cornell2020univpm,turpault2021sound},
more and more works tend to explore methods for the multiple events overlapping problem
in SED. Start from these two challenges, participants are encouraged to
propose systems that use sound separation jointly with the sound event detection.
Organizers provided a baseline SSep model~\cite{kavalerov2019universal, wisdom2020unsupervised} that can be
used for pre-processing to separate overlapping events and extract foreground
from the background sounds~\cite{turpault2020improving}.
Motivated by DCASE Task 4, \cite{huang2020guided} combined the SSep
and SED by fusing the general SED systems results with SSep-SED systems
that were trained using separated sources from an SSep system;
\cite{cornell2020univpm} proposed an SED-aware separation method
to remove the background noise rather than full foregrounds separation
from the mixtures.
Intuitionally, sound separation seems like a natural candidate to deal with
the overlapping SED~\cite{heittola2013supervised, kong2020source}.
However, there are still many problems when combing the SSep and SED,
for example, the separation system has its own error, which will
lead to wrong guidance of the SED system training, and if we joint train
the SSep and SED models, it is very difficult to balance multiple loss function between
these two tasks.
Therefore, how to effectively leverage the sound separation to improve
SED with overlapping acoustic events is very fundamental and challenging.

In this paper, we propose a new framework to utilize the sound separation
for reducing the impact of overlapping problem in SED. Two main contributions
are provided: 1) A novel selective pseudo-labeling (SPL) approach is proposed.
It produces high confidence single-event clips from the blind
sound separation outputs, by comparing their weakly labels with the corresponding
pseudo-labels that predicted by a well pre-trained audio tagging model.
These single-event clips are then used to improve the SED by
fine-tuning the original pre-trained model with a multi-objective learning strategy;
2) We provide a class-wise discriminative fusion to exploit the complementary
information between different SED models with mixture and
the SSep separated single-event recording inputs. This fusion is performed
at the frame-level predictions, using discriminative class-wise weights that
optimized on the SED development set.
All experiments are performed on DCASE 2021 Task 4 challenge,
the ``Sound Event Detection and Separation in Domestic Environments". Results
show that our proposed methods can achieve competitive performances with the
top ranked systems that reported in the DCASE challenge.

\section{Blind Sound Separation}
\label{sec:SSep baseline}

In DCASE 2021 Task 4 challenge, the official sound separation
and sound event detection baseline (we call it as \verb"SSep-J-MT" )
used a well pre-trained sound separation model to further improve
the pre-trained SED model \cite{Turpault2019_DCASE}.
This separation model follows a similar universal sound separation
system architecture as in \cite{kavalerov2019universal}, it can
separate an input sound mixture into a fixed number of sources.
The model is a TDCN++ masking network using STFT analysis/synthesis,
and it was trained in an unsupervised way with MixIT criterion
\cite{wisdom2020unsupervised} on the YFCC100m (1600 hours) dataset \cite{thomee2016yfcc100m}.
Using this well pre-trained SSep model, all the Task 4 data
was separated in an offline pre-processing way. Then,
the pre-trained SED model was fine-tuned on all of the
separated sound events. In \verb"SSep-J-MT" \cite{turpault2020improving},
predictions of target events were obtained by ensembling the
fine-tuned SED model with the original SED model using a learnt
weight during system training. The same pre-trained SSep model
in the \verb"SSep-J-MT" system is also used in our work.

\section{Proposed methods}
\label{sec:proposed}

Unlike using all of the separated sounds to fine-tune the
original SED model in the DCASE 2021 Task 4 \verb"SSep-J-MT" baseline, in this
work, we propose a new joint framework to combine the SSep and SED models
to enhance the overall sound event detection performance. The whole
framework is illustrated in Fig.\ref{fig:joint training framework}.
It contains four main modules, including the (a) offline SSep pre-processing,
(b) selective pseudo-labeling, (c) multi-objective model fine-tuning and (d)
class-wise discriminative score fusion.
Details of these modules are presented in the following sections.

\subsection{Selective Pseudo-labeling}
\label{sec:label-selection}

Detecting multiple overlapping sound events are typically more difficult
than isolated events. SSep can be used for SED by first separating the
single-event sounds in a mixed signal,
and then applying SED on each of them to improve the detection
performance~\cite{heittola2013supervised, kong2020source}.
Generally, the results obtained on well separated signals should be
more accurate than the ones on  original mixtures.
However, using SSep to improve SED is not a simple problem.
The separated sounds may contain errors or still a mixed signal,
they can also be background noise or non-target events. The
confidence of these separated signals are highly dependent
on how good the SSep model is. Therefore,
how to well exploit these separated signals during combing the SSep
and SED becomes interesting and important.

In this study, we aim to well utilize the available weakly and strong labeled mixtures
to improve the SED system, by choosing high-quality separated sounds which contain
only one single event that produced from an SSep system to update the original SED model.
Details are illustrated in Fig.\ref{fig:joint training framework}(b). Given an input mixture
signal $x$, it first be separated into $N$ clips, $x_{1}, x_{2}, ..., x_{N}$
by a well pre-trained SSep system,
TDCN++ \cite{kavalerov2019universal, tzinis2020improving}. Then, the clips are fed 
into a pre-trained audio tagging model (AT Model)
to predict their event class, $L=\{L_{1}, L_{2}, ..., L_{N}\}$, individually.
These predictions can be taken as pseudo-labels of the $N$ separated clips.
Next, we compare each $L_{i}$ with the available target event class label of $x$ and select those
high-quality ones according to the following criterion:
\begin{equation}
\label{scsc}
  S = \{(x_{i}, L_{i}) | L_{i} \in (G_{1}, G_{2}, ..., G_{k})\}, i \in 1,2,...,N
\end{equation}
where $S$ is the set of selected separated single-event clips whose
pseudo-labels  belong to the weak labels
of input mixture $x$. $G_{i}$
is the $i$-th weak label of $x$ and $k$ is the target event source number
that labeled for mixture $x$, such as, if $x$ contains cat, dog, dishes and speech, then $k=4$.
$N$ is 1 plus the maximum source number of each audio mixture in
the development set, because the input mixture may contain non-target events
or background noise that are taken as `other' class in our AT model.

Fig.\ref{fig:joint training framework}(e) demonstrates an example of the  pseudo-label
selection process (PL selection module in Fig.\ref{fig:joint training framework}(b)).
Providing each separated signal with pseudo-label,
we compare each pseudo-label with the ground-truth (collected from weak/strong label)
of event classes that
labeled for target sources in the mixture signal, if it
is single-event and can be found in the ground-truth, then we think that,
this single-event clip is separated well by  an SSep, and its pseudo-label
can be trust. As in Fig.\ref{fig:joint training framework}(e),
only the separated single-event clip with cat and speech are selected for
next multi-objective SED model fine-tuning. Actually, in our selective
pseudo-labeling, only the weak labels of training mixtures are required to
guarantee the quality of selected single-event pseudo-labels of the blind
sound separation outputs.

\subsection{Multi-objective Model Fine-tuning}
\label{sec:joint-training}

In the \verb"SSep-J-MT" baseline of DCASE 2021 Task 4~\cite{dcase2021web}, all the blind
separated sound clips are used to fine-tune the original sound event detection
model that trained on the mixtures with a multi-class network targets. However,
as discussed in the above section, the separation quality highly depends on the
SSep model, these separated sound clips may still contain many
 errors. The official \verb"SSep-J-MT" baseline may bring these errors
into the SED model fine-tuning and constrain the SSep potential for
improving the SED performance.

In this work, as the official \verb"SSep-J-MT", we still use the
same multi-objective model fine-tuning approach but different
inputs to improve the final SED model.
Details are shown in Fig.\ref{fig:joint training framework}(c).
Instead of using all the blind separated clips,  we only
take the high-quality single-event clips that selected according
to Equation (\ref{scsc}) to perform the SED model fine-tuning. In Fig.\ref{fig:joint training framework}(c),
\verb"Sum all" means directly adding all the selected single-event clips together
to form a new mixture sound. \verb"F-SED" represents the fine-tuned
SED model using selected single-event clips. \verb"pre "$\hat{N}$ means
the sound detection prediction of $\hat{N}$-th input clip.
With the pre-trained SED model as initialization,
the F-SED model is fine-tuned using a multi-objective
way to obtain the final prediction as follows:
\begin{equation}
\label{weight}
  {P_{final}} = \alpha  \cdot {P_{SED}} + \left( {1 - \alpha } \right) \cdot {P_{F - SED}}
\end{equation}
where $P_{final}$ is the final prediction of the joint module \verb"J-SED" (in 
Fig.\ref{fig:joint training framework}(c)) given a random
input sound mixture $x$, $P_{SED}$ and $P_{F - SED}$ represent the SED model and F-SED
model predictions respectively, and $\alpha$ is the weight parameter to be learnt.

Given the $P_{final}$ and freezing the pre-trained SED model,
the whole F-SED model is finally fine-tuned
using the standard total connectional binary cross entropy (BCE) loss
for weakly supervised teacher-student structure SED model training.
Details of training strategy can be found in our previous work \cite{liang2022joint}.

\subsection{Class-wise Discriminative Score Fusion}
\label{sec:score fusion}

Combining the sound event decisions of multiple SED models often can
result in better overall performance than using single model. In this
study, based on the provided SSep-SED joint framework,
we propose to combine multiple models at frame-level predictions,
using the weights that calculated to reflect
the class-wise sound event discrimination. The idea is motivated by
the weighted average of softmax form that used in ~\cite{salamon2017multiple, hayashi2020conformer}.
Given $M$ SED models, and frame-level posterior probability
$P_{cm}$ of class $c, c\in \{1,2,...,C\}$
on model $m, m\in\{1,2,...,M\}$, the final ensemble system
posterior probability of each frame belongs to class $c$, $P_{c}$ is defined as:
\begin{footnotesize}
\begin{equation}
\label{eq:softmax-fusion}
{P_c}\left( {{P_{cm}},F1_m^c} \right) = \frac{1}{M}\sum\limits_{m = 1}^M {{P_{cm}}\left[ {\frac{{\exp \left( {\beta  \cdot F1_m^c} \right)}}{{\sum\nolimits_{m' = 1}^M {\exp \left( {\beta  \cdot F1_{m'}^c} \right)} }}} \right]}
\end{equation}
\end{footnotesize}
where $F1_{m}^c$ is the development set class-wise $F1$-score \cite{mesaros2016metrics} of class $c$ on
model $m$ . $\beta$ is a tunable scalar parameter.

From Equation(\ref{eq:softmax-fusion}), it's clear that, the prediction
combination weight for each $P_{cm}$ can reflect the class-wise
importance of model $m$ for event class $c$, it means that if model $m$ can achieve
better $F1$-score on the development set, then, this model has a better
discrimination to class $c$, a larger weight will be assigned to $P_{cm}$ during
the multiple prediction combination.

The specific prediction combination is shown in
Fig.\ref{fig:joint training framework}(d). During SED inference,
for any input sound mixture $x$, we obtain three frame-level
predictions $P_{cm}$ and class-wise $F1$-scores using the joint module (J-SED),
the pre-trained SED and the F-SED model, respectively. The
former module using all the blind separated clips of $x$ as input,
the later two models directly take the raw mixture signal $x$
as input. The final prediction is achieved by combing the
three outputs using Equation(\ref{eq:softmax-fusion}).
As three SED models trained from different ways, the joint module
pays more attention to the well separated
single-event segments, while the original pre-trained model and
F-SED model may effectively avoid the errors introduced by
SSep. Therefore, we think that using class-wise
discrimination weights can well exploit the complementary information
between different models, by leveraging the individual advantage
of each model to produce better overall sound event detection performance.

\begin{figure}[t]
  \centering
  \includegraphics[width=7.5cm]{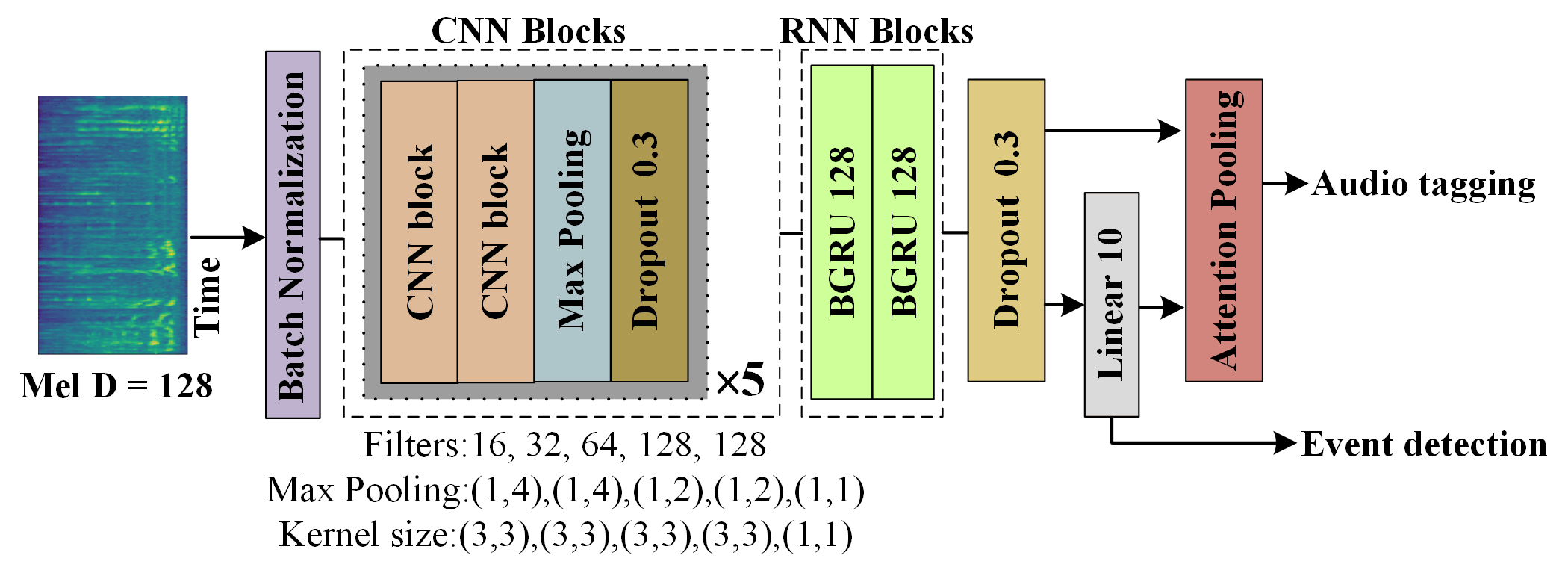}\\
  \caption{The student model structure of AT/SED/F-SED Model.}
  \label{fig:model}
\end{figure}

\section{Experiments and Results}
\label{sec:exp}

\subsection{Datasets and Features}
\label{sec:data-fea}

All our experiments are performed on the dataset of DCASE 2021 Task4 Challenge~\cite{turpault2020training}.
It is a sound event detection task in domestic environments.
The target of the systems is to provide not only the event class but
also the event time localization given that multiple events can be present
in an audio recording. The training set
includes 1,578 clips (2,244 class occurrences) of weakly labeled, 14,412 clips
of unlabeled and 10,000 strongly labeled audio clips. 1,168 strongly labeled
clips are taken as the develop validation clips. We extract 128 log mel-band
magnitudes as features, and each 10-second audio clip is transformed into 512 frames.

\subsection{Experimental setups}
\label{sec:setups}

We use the two DCASE 2021 Task 4 baseline systems \cite{turpault2020training} as our baselines,
one is the standard Mean-Teacher (\verb"MT") model \cite{delphin2019mean,
tarvainen2017mean}, and the other is the MT with sound separation (\verb"SSep-J-MT")
\cite{turpault2020improving}. Both \verb"MT "and \verb"SSep-J-MT "models have the same
teacher-student asymmetric neural network structure, and Fig.\ref{fig:model}
shows the details of their student branch (the teacher branch is exactly the same).
With the same model structure as \verb"MT", the \verb"AT", \verb"SED" and \verb"F-SED"
models in the joint framework of Fig.\ref{fig:joint training framework}
are improved by using our previous proposed adaptive focal loss training strategy
and event-specific post-processing technique during model training \cite{liang2022joint, Liang2021}.

The event-based $F1$-score (Collar-based F1) is used to measure the SED system performance.
It is computed with a 200ms collar on onsets
and a 200ms/20\% of the events length collar on offsets~\cite{mesaros2016metrics}.
In order to understand better what the behavior for different real scenarios
that emphasize different systems properties, DCASE 2021 Task 4 provided
two types of poly-phonic sound event detection scores (PSDS), PSDS1 and PSDS2 \cite{bilen2020framework}
as contrastive measures to evaluate systems.
$F1$-scores are computed using a single
operating point (decision thresholds=0.5) while PSDS values are computed
using 50 operating points (linearly distributed from 0.01 to 0.99).

\subsection{Results and Discussions}
\label{sec:result}

\subsubsection{Overall Results}
\label{sec:baselines result}

All our techniques are evaluated on the
DCASE 2021 Task 4 validation set,
because the ground-truth of test set is not released.
Results are shown in Table~\ref{tab:result-all}.
\verb"MT" and \verb"SSep-J-MT" are the official baselines.
\verb"SED", \verb"F-SED" and \verb"J-SED" represent three different sound event detection
models that shown in  Fig.\ref{fig:joint training framework}.
The detection inference procedures of system 2, 6 and 7 are shown in
Fig.\ref{fig:joint training framework}(d).
The prefix \verb"SSep" of system
1, 3 and 4-5 means the input mixture sounds are
first separated using the offline SSep pre-processing
(Fig.\ref{fig:joint training framework}(a)) before they
are fed into \verb"SED", \verb"F-SED" and \verb"J-SED" models during
inference, while systems 0, 2 and 7 without \verb"SSep" prefix means directly
using the original mixture sound as model inputs during SED inference.
System 3 is exactly the same as system 1 except for replacing the
\verb"MT" model by our \verb"SED", and it is also the same with
system 6 without our proposed selective pseudo-labeling block.
System 8, 9 and 10 are the three system fusion results using simple
equal weight average, logistic regression \cite{brummer2007focal} and
our proposed class-wise discriminative fusion that performed on
frame-level posterior probability of system 2, 6 and 7, respectively.

\begin{table}[th]
\caption{$F1$-scores (\%) and PSDS (\%) measures of the proposed methods.}
\label{tab:result-all}
\centering
\scalebox{0.84}{
\begin{tabular}{@{}clccc@{}}
\toprule
ID & System       & Collar-based F1 & PSDS1 & PSDS2 \\ \midrule
0  & \EscVerb{MT}~\cite{turpault2020training}           & 40.1            & 34.2  & 52.7  \\
1  & \EscVerb{SSep-J-MT}~\cite{turpault2020improving}      & 44.3            & 37.3  & 54.9  \\ \midrule
2  & \EscVerb{SED}~\cite{Liang2021}     & 42.0            & 41.8  & 71.7  \\
3  & \EscVerb{SSep-J-SED} (w/o SPL) & 43.0 & 42.0 & 72.1 \\ \midrule
4  & \EscVerb{SSep-SED}  & 42.8            & 41.8  & 71.9  \\
5  & \EscVerb{SSep-F-SED} & 43.6            & 42.1  & 72.2  \\
6  & \EscVerb{SSep-J-SED}  & 44.4            & 42.8  & 73.6  \\ \midrule
7  & \EscVerb{F-SED} & 42.6 & 42.1 & 74.1 \\ \midrule
8  & Avg(2,6,7) & 45.6 & 43.2 & 74.5 \\
9  & LR(2,6,7) & 45.7 & 42.9 & 74.6 \\
10  & Class-wise(2,6,7)  & 46.5 & 44.5 & 75.4 \\ \bottomrule
\end{tabular}}
\end{table}

From Table \ref{tab:result-all}, first,
 we see system 2, the \verb"SED" achieves significant performance
gains over system 0, \verb"MT", the collar-based F1 is improved from 40.1\% to 42.0\%,
and PSDS1, PSDS2 are improved from 34.2\%, 52.7\% to 41.8\% and
71.7\%. These big gains indicate that,
our previously proposed adaptive focal loss and ESP are
very effective to improve the baseline \verb"MT".

Second, comparing system 0 and 1, it's obvious to see the
effectiveness of sound separation to SED, and from
system 2 to 3, we also see performance gains
are brought by SSep with our \verb"SED" system, even these
gains are relatively small because of the improved \verb"SED".
This means that the SSep can reduce the impact of sound overlapping in SED to some extent.

Third, system 4, 5, 6 shows using
blind separated clips to test  three different models
in Fig.\ref{fig:joint training framework}(c). It's clear
to see that, the fine-tuned model, \verb"F-SED" is better than original
\verb"SED" model. And the joint module, \verb"J-SED " is better than
both  \verb"SED" and \verb"F-SED". It indicates that
the multi-objective model fine-tuning can effectively improve
the detection ability of SED system.
Because the final prediction of
\verb"J-SED " is the combination of \verb"SED" and \verb"F-SED"
with a learnt weight. Therefore, for SED inference with
SSep pre-processing, we only use system 6, \verb"SSep-J-SED" for
further system combination.

Fourth, by comparing system 6 and 3, it's clear that our proposed
selective pseudo-labeling
brings absolute 1.4\%, 0.8\% and 1.5\% collar-based F1, PSDS1 and PSDS2
improvement, respectively. And system 6 also outperforms the
official baseline-system 1 significantly in both PSDS1 and PSDS2.

Finally, to well exploit the complementary information in
all the three available SED models, the pre-trained \verb"SED",
the fine-tuned \verb"F-SED" and their joint model \verb"J-SED", we
choose to perform system fusion of 2, 6 and 7 to further improve
the final SED predictions, because they use the original input mixture
and SSep separated sounds in different ways with their best attempts.
Comparing results of system 8,9,10, we see the proposed class-wise discriminative
fusion not only performs better than other two traditional system fusion methods,
but also achieves much better performances than three single systems, either in
collar-based F1, or in the PSDS values.

\subsubsection{Class-wise Results}
\label{sec:classrst}

\begin{table}[th]
\caption{Class-wise performance (in collar-based $F1$-score (\%)) with different sound event detection systems.}
\label{tab:class-wise and fusion}
\centering
\scalebox{0.78}{
\begin{tabular}{@{}ccccc@{}}
\toprule
Event class/System         & \EscVerb{SED}      & \EscVerb{F-SED}   & \EscVerb{SSep-J-SED} & Class-wise(2,6,7)        \\ \midrule
Alarm/bell/ringing         & 38.2          & 38.3          & 41.1   & \textbf{42.3} \\
Blender                    & 43.3          & 43.2          & 44.6   & 44.8          \\
Cat                        & \textbf{69.2} & \textbf{69.4} & 68.7   & \textbf{71.0} \\
Dishes                     & 25.2          & 26.1          & 32.3   & \textbf{33.6} \\
Dog                        & 40.8          & \textbf{45.1} & 38.9   & \textbf{46.5} \\
Electric shaver/toothbrush & 17.3          & 17.4          & 19.9   & 20.3          \\
Frying                     & 45.8          & 45.0          & 48.9   & 49.2          \\
Running water              & 36.1          & 33.9          & 43.6   & 43.5          \\
Speech                     & 52.4          & 53.7          & 56.7   & \textbf{59.2} \\
Vacuum cleaner             & \textbf{51.7} & \textbf{53.9} & 49.3   & 54.1          \\ \midrule
Collar-based F1 (Average)  & 42.0          & 42.6          & 44.4   & 46.5          \\
\bottomrule
\end{tabular}}
\end{table}

Table \ref{tab:class-wise and fusion} presents the class-wise collar-based $F1$-scores
for system 2(\verb"SED"), 6(\verb"F-SED"), 7(\verb"SSep-J-SED") and their fusion results.
It's clear that different system has its own advantage to detect different
types of target events. Such as, \verb"SED" and \verb"F-SED" are better to detect
`Cat', `Dog' and `Vacuum cleaner', while \verb"SSep-J-SED "has more advantage
to detect other 7 types of events. Almost all the events detection
performances are improved by class-wise discriminative fusion,
especially the most obvious gains as shown in bold for `Alarm/bell/ringing', 
`Cat', `Dishes', `Dog' and `Speech', and the average $F1$ is also much better than
single systems. It means that the proposed fusion method can well exploit the 
detection ability of different models for different event classes.

\section{Conclusion}
\label{sec:conclution}

In this work, we propose a new training strategy of using sound separation
to enhance SED system. Different from previous works that directly using the blind
separated sound clips to fine-tune the SED model, we propose a selective pseudo-labeling (SPL)
method to select the high quality separated recordings for updating a joint SED
model, using the multi-objective model fine-tuning strategy.
In addition, a class-wise discriminative score fusion is further proposed, it is
performed at the frame-level posterior probabilities
to improve the final SED system performance. This score fusion uses the class-wise
weights to exploit each system's advantage on different sound event detection.
The proposed techniques are validated on the dataset of DCASE 2021 Task 4 Challenge,
experimental results show that both the proposed SPL and class-wise based score fusion
achieve significant performance improvements over the official baselines, and our
final SED system performances are competitive with the ones from
top ranked systems in DCASE 2021 Task 4 Challenge.
Our future work will focus on designing a better separation system to
improve the SED system.

\bibliographystyle{IEEEtran}

\bibliography{mybib}

\end{document}